\documentclass[aps,prl,twocolumn,superscriptaddress,showpacs,showkeys,amsmath,amssymb]{revtex4}
\usepackage{graphicx}% Include Figure files
\usepackage{dcolumn}% Align table columns on decimal point
\usepackage{bm}% bold math
\usepackage{color}
\usepackage[thinspace,thinqspace,amssymb,pstricks]{SIunits}
\usepackage{longtable}
\usepackage[letterpaper,hypertex]{hyperref}
\hypersetup{
    pdftitle={Is graphene on Ru(0001) a nanomesh?},    % title
    pdfauthor={Thomas Brugger, Sebastian G\"unther, Bin Wang, Hugo Dil, Marie-Laure Bocquet, J\"urg Osteralder, Joost Wintterlin, Thomas Greber},     % author
    pdfsubject={Comparison of the electronic structures of graphene on Ru(0001) and h-BN on Ru(0001).},   % subject of the document
    pdfkeywords={graphene; nanomesh; ruthenium; xenon; photoemission; thermal desorption spectroscopy}, % list of keywords
}

\begin{document}

\title{Is graphene on Ru(0001) a nanomesh?}

\date{\today}

%\author{T. Brugger$^1$, S. G\"unther$^2$, B. Wang$^2$, H. Dil$^{1,3}$, M.-L. Bocquet$^2$, J. Osterwalder$^1$, J. Wintterlin$^2$, and T. Greber$^1$}
%\affiliation{$^1$ Physik-Institut, Universit\"{a}t Zurich-Irchel, Winterthurerstrasse 190, CH-8057 Z\"{u}rich, Switzerland}
%\affiliation{$^2$ LMU, Germany}
%\affiliation{$^3$ PSI, Switzerland}

\author{Thomas Brugger}
\affiliation{Physik-Institut, Universit\"{a}t Z\"{u}rich, Winterthurerstrasse 190, CH-8057 Z\"{u}rich, Switzerland}
\author{Sebastian G\"unther}
\affiliation{Department Chemie, Ludwig-Maximilian Universit\"{a}t, Butenandtstrasse 5-13, D-81377 M\"{u}nchen, Germany}
\author{Bin Wang}
\affiliation{Universit\'e de Lyon, Laboratoire de Chimie, \'Ecole Normale Sup\'erieure de Lyon, CNRS, France}
\author{Hugo Dil}
\affiliation{Physik-Institut, Universit\"{a}t Z\"{u}rich, Winterthurerstrasse 190, CH-8057 Z\"{u}rich, Switzerland}
\affiliation{Swiss Light Source, Paul Scherrer Institute, CH-5232 Villigen, Switzerland}
\author{Marie-Laure Bocquet}
\affiliation{Universit\'e de Lyon, Laboratoire de Chimie, \'Ecole Normale Sup\'erieure de Lyon, CNRS, France}
\affiliation{Department Chemie, Ludwig-Maximilian Universit\"{a}t, Butenandtstrasse 5-13, D-81377 M\"{u}nchen, Germany}
\author{J\"urg Osterwalder}
\affiliation{Physik-Institut, Universit\"{a}t Z\"{u}rich, Winterthurerstrasse 190, CH-8057 Z\"{u}rich, Switzerland}
\author{Joost Wintterlin}
\affiliation{Department Chemie, Ludwig-Maximilian Universit\"{a}t, Butenandtstrasse 5-13, D-81377 M\"{u}nchen, Germany}
\author{Thomas Greber}
\email{greber@physik.uzh.ch}
\affiliation{Physik-Institut, Universit\"{a}t Z\"{u}rich, Winterthurerstrasse 190, CH-8057 Z\"{u}rich, Switzerland}

\begin{abstract}
The electronic structure of a single layer graphene on Ru(0001) is compared with that of a single layer hexagonal boron nitride nanomesh on Ru(0001). Both are corrugated sp$^2$ networks and display a $\pi$-band gap at the $\overline{\text{K}}$ point of their $1\times1$ Brillouin zone. Graphene has a distinct Fermi surface which indicates that 0.1 electrons are transferred per $1\times1$ unit cell. Photoemission from adsorbed xenon identifies two distinct Xe 5p$_{1/2}$ lines, separated by 240 meV, which reveals a corrugated electrostatic potential energy surface. These two Xe species are related to the topography of the system and have different desorption energies.
\end{abstract}



\maketitle

%\linenumbers

%\section{INTRODUCTION}

A single layer of an adsorbate strongly influences the physical and chemical properties of a surface. Sticking and bonding of atoms and molecules may change by orders of magnitude as well as the transport properties across and parallel to the interface. For developments in nanotechnology it is particularly useful to have single layer systems which are inert, remain clean at ambient conditions and are stable up to high temperatures. In this field sp$^2$ hybridized graphene and hexagonal boron nitride nanomesh are outstanding examples \cite{nov04,cor04}. On ruthenium both form perfect single layers, where the lattice mismatch between the substrate and the adsorbate causes two dimensional regular super structures with a lattice constant of about $\unit{3}{\nano\meter}$ \cite{mar07,pan07,par08,gor07}. A nanomesh is a corrugated single layer dielectric. In the case of the $h$-BN/Rh(111) nanomesh, that has an atomic and electronic structure like $h$-BN/Ru(0001), single molecules separated by $\unit{3}{\nano\meter}$ are observed after adsorption at room temperature \cite{ber07}. For graphene on Ru (g/Ru(0001)) similar, but also complementary properties are expected.

\begin{figure}
\includegraphics[width=0.88\columnwidth]{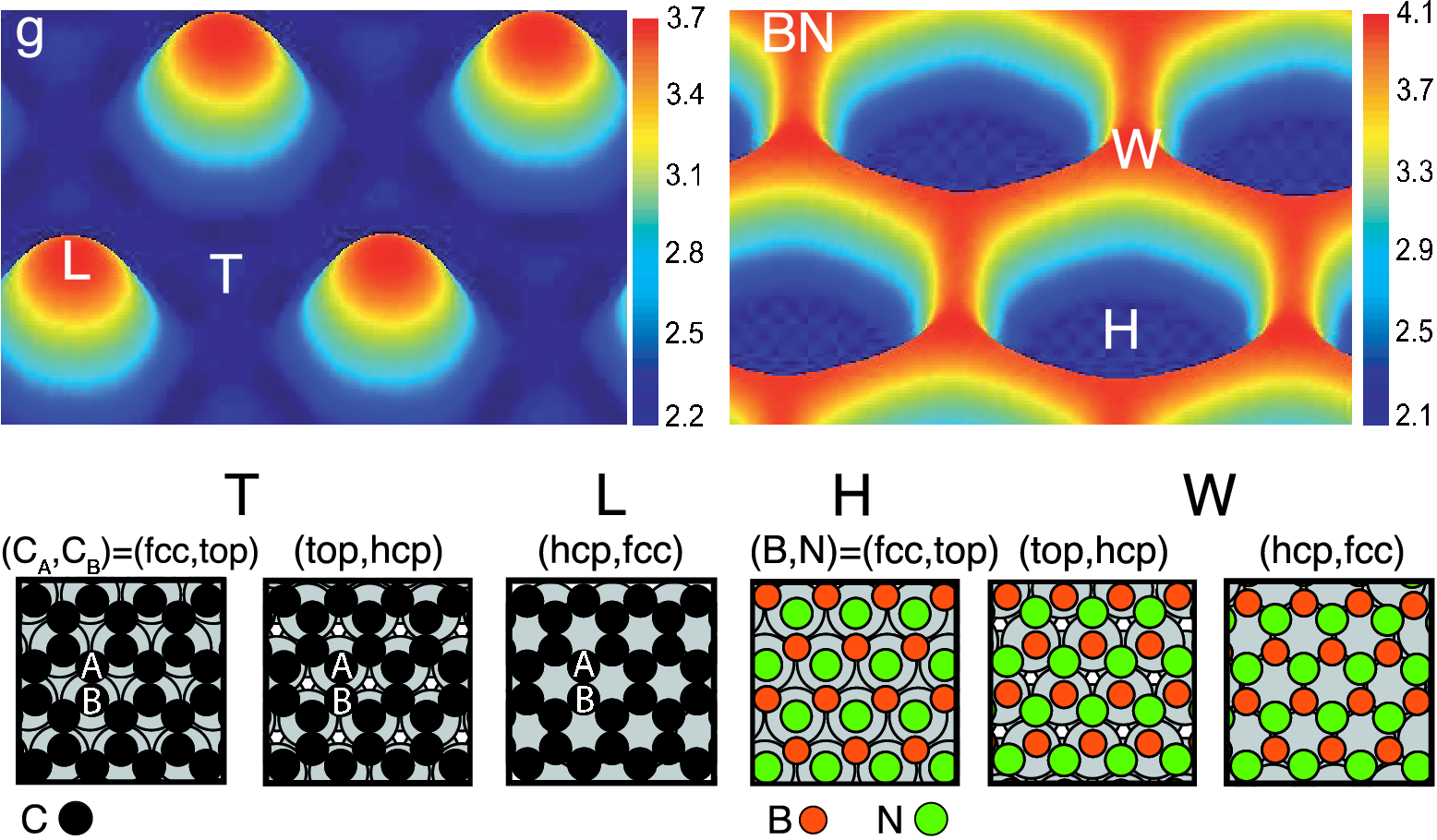}
\caption{\label{F1}(Color online) Views of the height (\AA) modulated graphene (g) and $h$-BN nanomesh (BN) on Ru(0001), as obtained from a DFT calculation for both systems. L and T  denote loosely and tightly bound graphene, H and W holes and wires of the $h$-BN nanomesh. The lower six panels illustrate the three different regions ((fcc,top), (top,hcp) and (hcp,fcc)) which can be distinguished in both systems (see text).}
\end{figure}
The purpose of this letter is to establish a comparison between g/Ru(0001) and $h$-BN/Ru(0001) using photoemission and Density Functional Theory (DFT). It is shown that g/Ru(0001) is a metal with a sizeable Fermi surface, while $h$-BN/Ru(0001) is not. Though, the exploration of the electrostatic potential energy landscape by photoemission of adsorbed Xe indicates also a modulation of the local workfunction for g/Ru(0001), analog to $h$-BN/Rh(111) \cite{dil08}.

The atomic structure of graphene and boron nitride sp$^2$ networks on transition metals is driven by coincidence or incommensurate lattices (see Figure \ref{F1}) \cite{lan92,ndi06,mar07,pan07,par08,paf90,cor04,mor06,pre07b}.
\begin{figure*}
\includegraphics[width=0.387\textwidth]{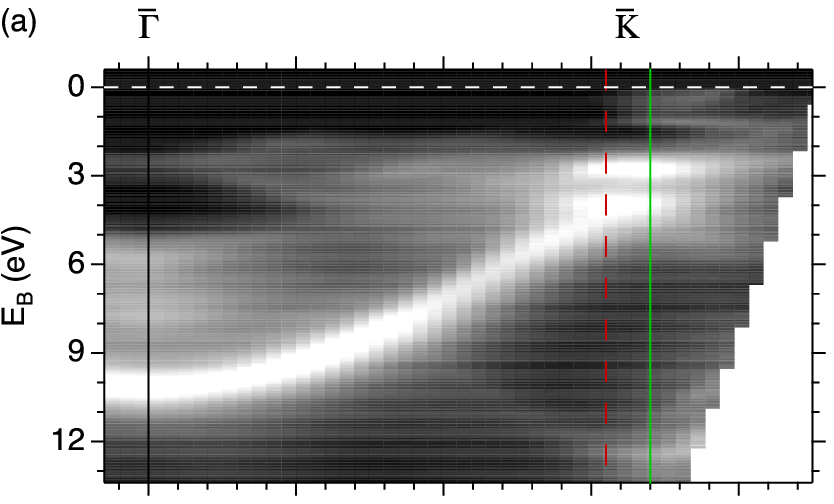}
\includegraphics[width=0.379\textwidth]{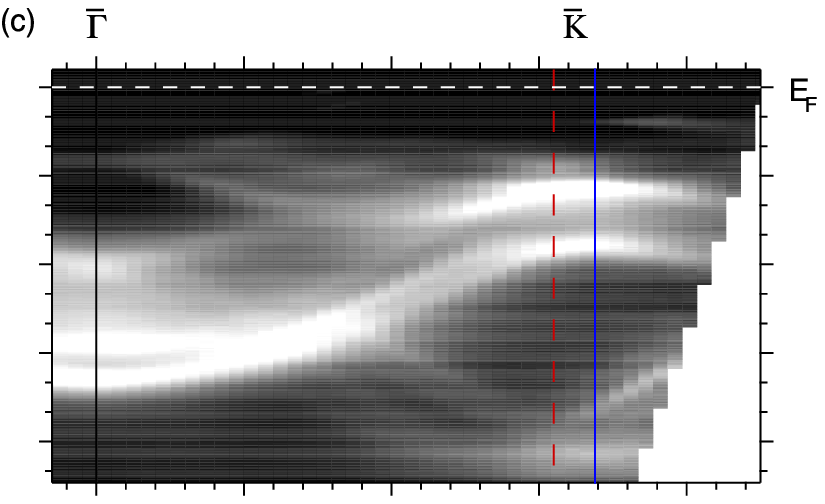}

\includegraphics[width=0.387\textwidth]{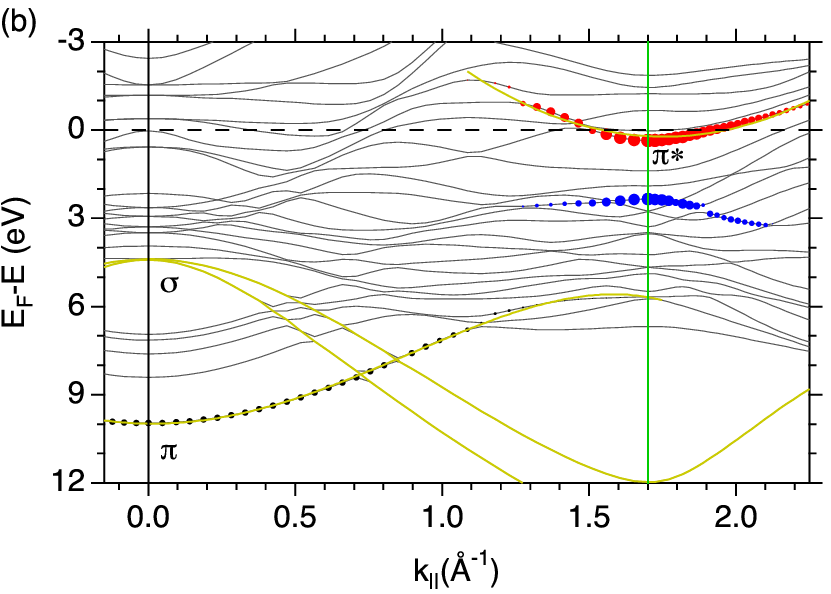}
\includegraphics[width=0.379\textwidth]{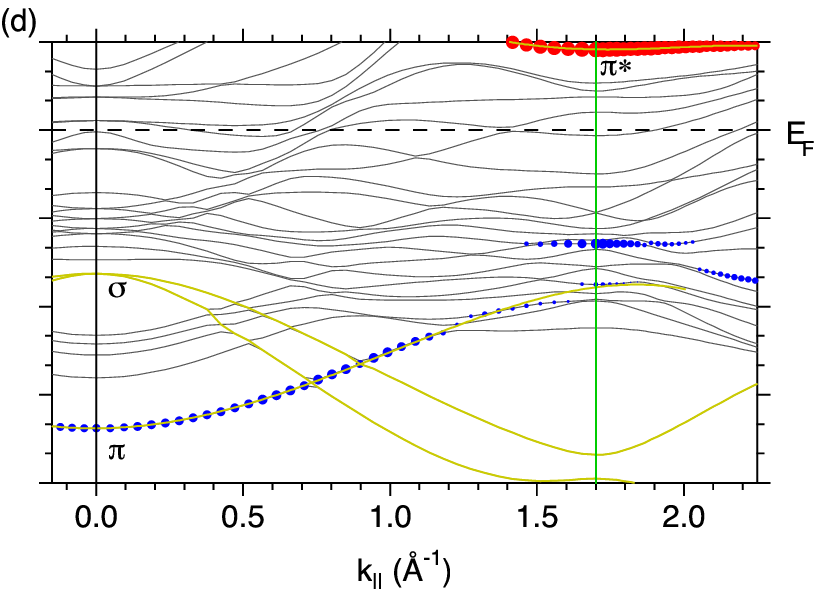}
\caption{\label{F2} (Color online) Bandstructures of graphene and $h$-BN nanomesh on Ru(0001) along $\overline{\Gamma}~\overline{\text{K}}$. (a) He II$_\alpha$ photoemission of g/Ru(0001). (b) DFT of g/Ru(0001) for (C$_\text{A}$,C$_\text{B}$)$\sim$(top,hcp). (c) He II$_\alpha$ photoemission of $h$-BN/Ru(0001). (d) DFT of $h$-BN/Ru(0001) for (B,N)$\sim$(fcc,top). The vertical lines at $\overline{\text{K}}$ indicate the boundaries of the $1\times1$ surface Brillouin zones for Ru (red dashed), $h$-BN (blue solid) and graphene (green solid). The size of the filled circles in (b) and (d) represents the $p_z$ weight of the adsorbate atoms on the bands, where blue describes top atoms (C$_\text{A}$ in (b) and N in (d)) and red hollow site atoms (C$_\text{B}$ in (b) and B in (d)). Black circles depict the average of top and hollow site atoms. Thick yellow curves are guides for the eyes.}
\end{figure*}
For $h$-BN/Rh(111) $13\times13$ BN units coincide with $12\times12$ Rh units \cite{cor04,bun07} and similar lattice constants have been observed for adsorbed graphene systems \cite{par08,mar07,ndi06}. The concomitant variation of the local coordination of the substrate and the adsorbate atoms divides the unit cells into regions with different lateral coordination. The notation (B,N)$\sim$(top,hcp) refers to the local configuration, where a B atom sits on top of the Ru atom in the first substrate layer and N on top of the hexagonal close packed (hcp) site, i.e. on top of the Ru atom in the second layer. Therefore 3 regions (fcc,top), (top,hcp) and (hcp,fcc) can be distinguished (see Figure \ref{F1}). Whereas BN has a base with two different atoms in the unit cell, the base of graphene consists of two identical carbon atoms C$_\text{A}$ and C$_\text{B}$ which become distinguishable by the local coordination to the substrate. The local coordination of the (B,N) or (C$_\text{A}$,C$_\text{B}$) units in the honeycomb overlayer gradually shifts along the diagonal of the unit cell from (hcp,fcc) via (fcc,top) and (top,hcp) back to (hcp,fcc). In g/Ru the local (fcc,top) and (top,hcp) coordination leads to close contact between the (C$_\text{A}$,C$_\text{B}$) atoms and the substrate \cite{wan08} while (B,N) is strongly interacting only in the (fcc,top) coordination \cite{las08}. As a result, twice as many atoms are bound in strongly interacting regions in g/Ru when compared to $h$-BN/Ru. In the following we call the tightly bound region $T$-region and the loosely bound [(C$_\text{A}$,C$_\text{B}$)$\sim$(hcp,fcc)] $L$-region. This causes the most obvious difference in the atomic structure of the two sp$^2$ networks.

The single layer graphene has been grown in ultrahigh-vacuum by thermal decomposition of $\unit{30}{L}$ ($\unit{1}{L}=\unit{10^{-6}}{Torr\ \second}$) ethene (C$_2$H$_4$) on the $\unit{1100}{\kelvin}$ hot Ru(0001) surface which had been cleaned by standard procedures. It was characterized with scanning tunneling microscopy and low energy electron diffraction.

\textit{Ab initio} calculations are performed with the VASP package based on DFT, which implements PAW pseudopotentials \cite{kre99} and the PBE exchange correlation functional in the Generalized Gradient Approximation (GGA) \cite{per96}. For the band structure calculations epitaxial ($1\times1$) C or BN/Ru structures are investigated with the lattice constant of graphene, a 4-layer Ru slab,  a $36\times36\times1$ $k$-sampling and $\unit{400}{\electronvolt}$ cutoff. For the calculation of the atomic structure and the electrostatic potential, large moir\'e periodicities are considered with asymmetric (structure) and symmetric (potential) slabs.

The hybridization of the carbon p$_z$ orbitals with the substrate atoms breaks the symmetry between the C$_\text{A}$ and the C$_\text{B}$ atoms. This is reflected in the band structure where a large $\pi$-band gap opens at $\overline{\text{K}}$. Figure \ref{F2}(a) shows the measured band structure for g/Ru(0001) along $\overline{\Gamma}~\overline{\text{K}}$. At $\overline{\text{K}}$ the $\pi$-band levels off at a binding energy of $\unit{4.6\pm0.1}{\electronvolt}$. This strong hybridization is in line with observations on g/Ni(111) \cite{nag95a}. The experiment is in good agreement with calculations for a ($1\times1$) graphene sheet $\unit{2.2}{\angstrom}$ above the topmost Ru layer with (C$_\text{A}$,C$_\text{B}$)$\sim$(top,hcp) (Figure \ref{F2}(b)). Figure \ref{F2}(c) shows the same section of $k$-space for $h$-BN/Ru(0001). The $\pi$-band levels off at a binding energy of $\unit{5.4\pm0.1}{\electronvolt}$. Also $h$-BN on Ru is well described with calculations for a ($1\times1$) (B,N)$\sim$(fcc,top) sheet $\unit{2.2}{\angstrom}$ above the topmost Ru layer (Figure \ref{F2}(d)). Theory also shows that the two atoms in the base of the sp$^2$ networks have a different $p_z$ weight on the different bands. For the case of BN the $\pi$-band is mainly nitrogen derived, while the unoccupied $\pi^*$-band has its main weight on the boron atoms. For the ($1\times1$) model of graphene only the C$_\text{B}$ atom contributes to the $\pi^*$-band while the $\pi$-band has equal portions from top and hollow site atoms. The effect of $p_z-d$ C-Ru hybridization is reflected by localized states in the $\pi$-bandgap that are mainly C$_\text{A}$-Ru derived.
%hier muessen wir nochmals genau ueberpruefen, ob das stimmt, die C_A C_B vertauschung muss ueberall ueberprueft werden.
A distinct difference between graphene and $h$-BN nanomesh is the $\sigma$-band splitting of $\unit{0.9}{\electronvolt}$ for the BN bands that belong to the `wires' and the `holes' respectively \cite{gor07}. The band splitting of $h$-BN was assigned to the dielectric nature of $h$-BN and the workfunction difference between the `hole' and the `wire' regions. For graphene no splitting is observed, where a splitting smaller than $\unit{310}{\milli\electronvolt}$ could not be resolved by the experiment. This can be explained by a smaller corrugation of the graphene layer or by the metallic nature of graphene.

\begin{figure}
\includegraphics[width=0.46\columnwidth]{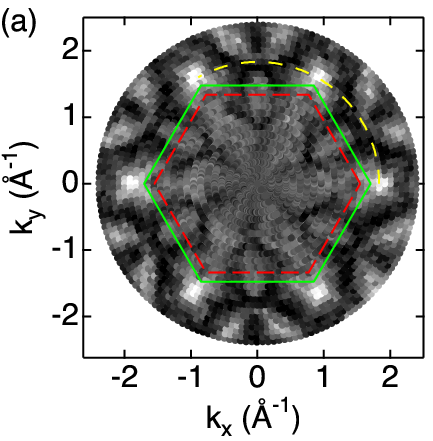}
\includegraphics[width=0.412\columnwidth]{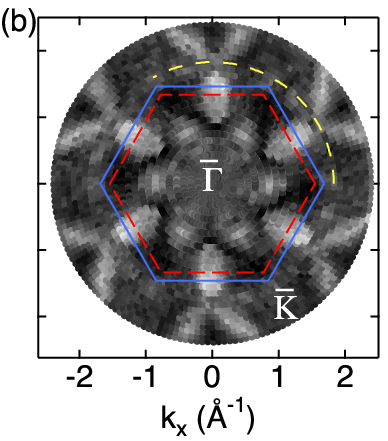}

\includegraphics[width=0.459\columnwidth]{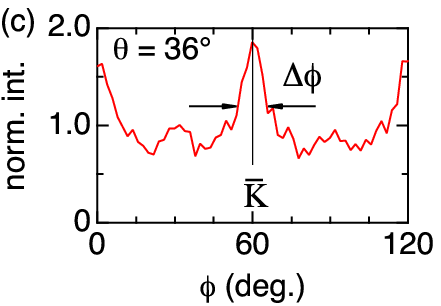}
\includegraphics[width=0.397\columnwidth]{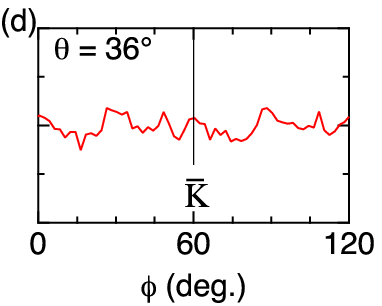}
\caption{\label{F3} (Color online) He II$_\alpha$ Fermi surface maps. (a) g/Ru(0001). (b) $h$-BN/Ru(0001). The hexagons indicate the surface Brillouin zones of Ru(0001) (red dashed), graphite (green solid) and $h$-BN (blue solid). (c) and (d) show the normalized intensities of azimuthal cuts along the dashed yellow sectors in (a) and (b) respectively.}
\end{figure}

The metallicity of g/Ru(0001) is reflected in the measured Fermi surface map (FSM) which is compared to that of $h$-BN/Ru(0001) in Figure \ref{F3}.
The FSM of $h$-BN/Ru(0001) shows only bands that are also seen on the bare Ru(0001) surface. On the other hand, g/Ru(0001) displays states at the Fermi level that are reminiscent to the Dirac points at the $\overline{\text{K}}$ points of free standing graphene \cite{bos07a}. The bandstructure measurements in Figure \ref{F2}(a) demonstrate that these graphene related states are part of the $\pi^*$-band, which means that charge is transferred from the substrate to the graphene. The Luttinger volume of the electron pockets near $\overline{\text{K}}$ corresponds to the number of transferred electrons  $N_e=2\pi/\left(3\sqrt{3}\right)\left(\Delta\phi_\text{K}\right)^2$,
%\begin{equation}
%N_e=\frac{2\pi}{3\sqrt{3}}\left(\Delta\phi_\text{K}\right)^2
%\end{equation}
where $\Delta\phi_\text{K}$ is taken as the full width at half maximum of the intensity on an azimuthal cut across $\overline{\text{K}}$, in radians (Figure \ref{F3}(c)). For the FSM in Figure \ref{F3}(a) we find $\Delta\phi_\text{K}=\unit{0.27\pm0.03}{\radian}$, which translates to $N_e=0.09\pm0.02$ electrons per $1\times1$ graphene unit cell. Of course, this value is the average for the whole graphene unit cell.

\begin{figure}
\includegraphics[width=0.79\columnwidth]{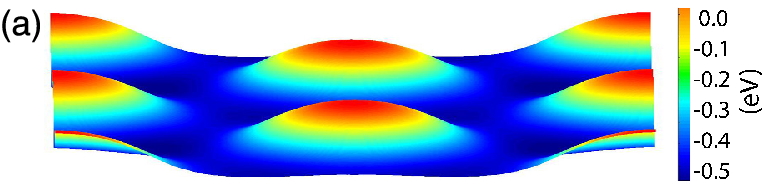}

\includegraphics[width=0.389\columnwidth]{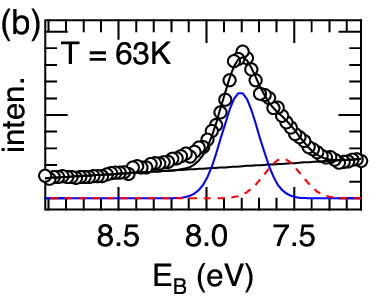}
\includegraphics[width=0.389\columnwidth]{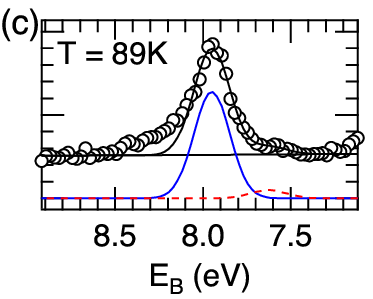}

\includegraphics[width=0.79\columnwidth]{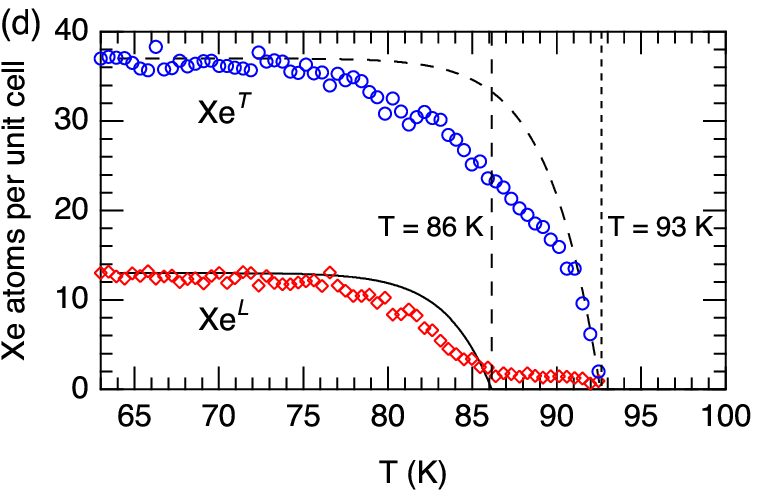}
\caption{\label{F4} (Color online) Xe adsorption on g/Ru(0001). (a) Calculated electrostatic potential map $\unit{3.8}{\angstrom}$ above the outmost carbon atom. (b) and (c) He I$_\alpha$ excited Xe 5p$_{1/2}$ spectra for two Xe coverages during thermal desorption. At high coverage two spectral components may be distinguished, as can be seen from the two Gaussians that are fitted to the data. (d) Spectral weights of the two Xe species as a function of temperature. Blue open circles stand for the high binding energy and red open diamonds for the low binding energy component.}
\end{figure}

The differences in the atomic and electronic structure are also reflected in the potential energy surfaces that drive the functionality of the super structures as templates for the formation of molecular arrays.
In Figure \ref{F4}(a) the calculated electrostatic potential for g/Ru(0001) $\unit{3.8}{\angstrom}$ above the carbon atoms is shown.
As for the case of $h$-BN/Rh(111) nanomesh the energy corrugation correlates with the atomic corrugation \cite{dil08}. This potential is measured with photoemission of adsorbed xenon \cite{wan84a}. Also for g/Ru(0001) two Xe bonding regions can be distinguished with distinct Xe 5p$_{1/2}$ photoemission binding energies and spectral weight (Figure \ref{F4}(b)). Assuming a Xe van der Waals radius of $\unit{2.2}{\angstrom}$, 50 Xe atoms per g/Ru(0001) unit cell are expected for the monolayer coverage \cite{mar08}. From the spectral weight and the known atomic structure it can be deduced that the Xe$^T$ species on the tightly  bonded graphene has the higher Xe 5p$_{1/2}$ photoemission binding energy and contains 37 atoms at full coverage. The Xe$^L$ species with lower photoemission binding energy corresponds to Xe adsorbed on the loosely bonded graphene. The binding energy difference between these two species of $\unit{236\pm5}{\milli\electronvolt}$ is determined from a fit of two Gaussians with equal width and does not depend on the coverage. This value reflects a local work function difference between $T$- and $L$-graphene and is in agreement with the theoretical result shown in Figure \ref{F4}(a). Like on the $h$-BN nanomesh the tightly bonded regions have a lower work function. However, for g/Ru(0001) the local work function difference is about $\unit{25}{\%}$ lower than for the case of $h$-BN/Rh(111). Intuitively, this is related to the metallic nature of the graphene that screens out lateral electric fields in the graphene. The better screening of graphene is indeed reflected in the Xe 5p$_{1/2}$  final state binding energy as referred to the vacuum level $E_B^V=E_B+\Phi$ \cite{kai80}. Table \ref{T1} shows that $E_B^V$ of Xe$^T$ on g/Ru(0001) is 210 meV smaller than that of Xe$^H$ on $h$-BN/Rh(111).

From the thermal desorption the Xe adsorption energy is inferred. In Figure \ref{F4}(d) the spectral weights of the Xe species are shown as a function of temperature (heating rate $\beta=\unit{1.5\pm 0.05}{\kelvin/\minute}$). The temperatures at which the two Xe species disappear indicate that Xe$^T$ is about $\unit{8}{\%}$ stronger bound than Xe$^L$. In order to compare g/Ru with $h$-BN/Rh \cite{dil08}, the temperature dependent weights of the two Xe species were fitted to zero order desorption. From $-dN=\nu/\beta\exp(-E_d/k_\text{B}T)dT$ the desorption energies $E_d$ are found (see Table \ref{T1}).
\begin{table}
\begin{tabular}{|l|r|r|r|r|r|}
\hline
&\multicolumn{3}{c|}{$h$-BN/Rh(111)}&\multicolumn{2}{c|}{g/Ru(0001)}\\
\hline
Phase&$C^W$&$C^H$&$R^H$&Xe$^L$&Xe$^T$\\
\hline
$E_d\unit{}{\left(\milli\electronvolt\right)}$&$181$&$184$&$208$&$222$&$231$\\
$N_1$&25&17&12&13&37\\
\cline{2-6}
$E_B\unit{}{\left(\electronvolt\right)}$&$7.42$&\multicolumn{2}{r|}{$7.72$}&$7.56$&$7.80$\\
\hline
$\Phi\unit{}{\left(\electronvolt\right)}$&\multicolumn{3}{r|}{$4.18$}&\multicolumn{2}{r|}{$3.89$}\\
\hline
\end{tabular}
\caption{\label{T1} Experimentally determined Xe parameters for $h$-BN/Rh(111) \cite{dil08} and g/Ru(0001): desorption energies $E_d$, number of Xe atoms at full coverage $N_1$. For all fits an attempt frequency $\nu$ of $\unit{1.2\times10^{12}}{\hertz}$ has been used. Xe 5p$_{1/2}$ photoemission binding energies $E_B$ and global workfunctions $\Phi$ for monolayer coverage. The errors for the binding energies $E_B$ and the workfunctions $\Phi$ are $\unit{\pm0.02}{\electronvolt}$ and  $\unit{\pm2}{\milli\electronvolt}$ for the desorption energies $E_d$.}
\end{table}
The values are slightly smaller than the desorption energies on graphite ($\unit{249}{\milli\electronvolt}$ \cite{ulb06}) and higher with respect to $h$-BN/Rh. The fits for zero order desorption show that the Xe$^T$ species are not well described with a single desorption energy (dashed line in Figure \ref{F4}(d)). In the $h$-BN/Rh case two Xe$^H$ phases (C$^H$ and R$^H$) had been identified by a pronounced kink in the desorption spectrum of Xe$^H$. The more strongly bound of these two phases was assigned to Xe$^H$ atoms at the rims of the holes where dipole rings induce an enhanced polarization and bonding. For g/Ru(0001) a clear kink is not visible, but the deviations from the fit also indicate variations in bonding strength of the Xe$^T$ atoms. The difference to $h$-BN/Rh may be understood by the different shape of the potential energy surface. If one nevertheless fits two Xe$^T$ phases binding energies of $\unit{222}{\milli\electronvolt}$ and $\unit{234}{\milli\electronvolt}$ are obtained for g/Ru(0001). The small difference of $\unit{12}{\milli\electronvolt}$ is not unreasonable considering the lower work function modulation than for $h$-BN/Rh.

In conclusion, the presented findings suggest that g/Ru is not a nanomesh, i.e. a corrugated single layer dielectric, but a corrugated single layer metal. However, it is also expected to act as a template for molecular trapping, where the metallicity of graphene imposes a stronger electronic coupling of adsorbates to the underlying transition metal.

Fruitful discussions with Peter Blaha and technical support by Martin Kl\"ockner are gratefully acknowledged. We thank the Swiss National Science Foundation and the German Research Foundation for their financial support. M.L.B thanks the Humboldt foundation for a research fellowship. Part of this work was performed at the Swiss Light Source.

\end{document}